\documentclass{aa}
\usepackage{psfig}

\newcommand{\cm}{cm$^{-2}$}

\newcommand{\dla}{ damped Lyman-$\alpha$ }
\newcommand{\beq}{\begin{equation}}
\newcommand{\eeq}{\end{equation}}
\newcommand{\noi}{\noindent}

\newcommand{\dV}{\Delta V}

\begin{document}
\thesaurus{11(11.05.2; 11.06.1; 11.09.4), 12(12.03.3), 13(13.19.3)}

\title{HI 21~cm absorption in low $z$ damped Lyman-$\alpha$ systems}
\titlerunning{21~cm from low $z$ DLAS}

\author{Nissim Kanekar \thanks{nissim@ncra.tifr.res.in}, 
Jayaram N Chengalur    \thanks{chengalur@ncra.tifr.res.in}}
\authorrunning{Kanekar \& Chengalur }
\institute{National Centre for Radio Astrophysics, Post Bag 3, Ganeshkhind, Pune 411 007}
\date{Received mmddyy/ accepted mmddyy}
\offprints{Nissim Kanekar}
\maketitle

\begin{abstract}
We report Giant Metrewave Radio Telescope (GMRT) 21~cm observations of two confirmed
and one candidate low redshift damped Lyman-$\alpha$ systems (DLAS). HI absorption was 
detected in the two confirmed systems, at $z = 0.2378$ and $z = 0.5247$, with spin 
temperatures of $2050 \pm 200$~K and $330 \pm 70$ K respectively. The non-detection of 
absorption in the third (candidate) system places an upper limit on its 
optical depth, which in turn implies a $3\sigma$ lower limit of 850~K on its spin 
temperature. Ground based K-band imaging observations (\cite{raovla}) have 
identified the low ${\rm T_s}$ absorber with a spiral galaxy while the high 
${\rm T_s}$ system appears to be an LSB galaxy. This continues the trend 
noted earlier by Chengalur \& Kanekar (2000) of low spin temperatures  
for DLAS associated with spiral galaxies  and high spin temperatures for 
those  with no such association.

	Combining these new results  with 21 cm observations from the published 
literature, we find that (1)~The 21 cm equivalent width ($EW_{21}$) of DLAS 
increases with increasing $N_{\rm {HI}}$. Further, absorbers associated with spiral 
galaxies have systematically higher $EW_{21}$ at the same $N_{\rm HI}$ than 
DLAS with no such  association, (2)~DLAS at high redshift ($z \ga 2$) have 
small velocity spreads $\Delta V$, while, at low redshift ($z \la 2$), both 
large and small $\Delta V$ are observed. These trends are in qualitative agreement 
with hierarchical models of galaxy formation.
\end{abstract}

\section{Introduction}
\label{sec:intro}

Damped Lyman-$\alpha$ systems (DLAS), are the rare, high HI column density 
($N_{\rm HI} \ga 10^{20}$ cm$^{-2}$) absorbers seen in spectra taken against distant 
quasars. These systems are the major observed repository of neutral gas at high 
redshift ($z \sim 3$) and are thus logical candidates for the precursors of modern-day 
spiral galaxies (\cite{wolfe88}). However, despite more than a decade of systematic 
study, the 
structure and evolution of the absorbers remains an issue of much controversy. The original 
optical surveys to detect DLAS were carried out with the prime motivation of 
detecting disk galaxies at high redshift (e.g. Wolfe et al. 1986). Follow-up 
observations of the kinematics of the absorbers (as revealed by their unsaturated, 
low ionization metal line profiles) showed that these profiles are asymmetric in 
most DLAS. The asymmetries can be simply interpreted if the absorption arises 
in a thick, rapidly rotating disk (\cite{pw1,pw2}), 
consistent with the idea that the absorbers are the progenitors of massive spiral 
galaxies.  However, it has been shown that the observed line profiles can also be 
explained by mergers of sub-galactic clumps in hierarchical clustering scenarios 
(\cite{haehnelt98}), by dwarf galaxy ejecta (\cite{nulsen98}) and even by randomly moving 
clouds in a spherical halo (\cite{mcdonald99}). Further, Ledoux et al. (1998)
found that the asymmetry of the metal line profiles is pronounced only for systems
with $\Delta V < 150$ km s$^{-1}$; this is contrary to what would be expected
if DLAS were indeed massive rotating disks (see, however, Wolfe \& Prochaska 1998).

DLAS are known to contain some dust (Pei, Fall \& McMahon 1989) and the problems of 
quantitatively accounting for dust depletion make the study of their metallicities and chemical 
evolution difficult. Studies using the abundances of metals like
S or Zn, which are only slightly depleted onto dust grains (\cite{pettini97,pettini99}), 
as well as attempts at modelling dust depletion (\cite{vladilo98}), indicate that the 
absorbers have low metallicities ($Z \la 0.1 Z_\odot$) and do not show much metallicity 
evolution with redshift. Further, DLAS do not show the [$\alpha$/Fe] 
enrichment pattern that characterizes  low metallicity halo stars in the Milky Way 
(Pettini et al. 1999, Molaro et al. 1998, Centuri\'{o}n et al. 2000). This difference 
in the [$\alpha$/Fe] enrichment pattern suggests that DLAS have star formation histories 
different from spirals and more like those of dwarf galaxies.

In the local universe, the overwhelming contribution to the cross section for DLA
absorption is believed to be made by the disks of spiral galaxies 
(\cite{rb93,rt98}); one might then expect at least the low redshift DLAS to 
be associated with spiral disks. Hubble Space Telescope (HST)and 
ground-based imaging have, however, shown that low~$z$ DLAS are, in fact, 
associated with a wide variety of morphological types (\cite{lebrun97,rt98,raovla})
and are not exclusively (or even predominantly) large spirals.

If the QSO behind the absorbing system is radio-loud, 21~cm absorption 
observations can be used to probe physical conditions and kinematics in the 
absorbers (see, for example, \cite{ck2000,lane2000,carilli}). Such studies 
directly yield the spin temperature of the absorbing gas, which, for a homogeneous
absorber, is typically the same as its kinetic temperature (\cite{field1958})
For a multi-phase absorber, the measured spin temperature is the
column density weighted harmonic mean of the temperatures of the individual phases.  

Chengalur \& Kanekar (2000) found that low ${\rm T_s}$ values (${\rm T_s} \la 300$ K) were obtained 
in the few cases where the absorber was identified to be a spiral galaxy. Such 
temperatures are typical of the Milky Way and local spirals (\cite{bw1992,braun1997}). 
However, the majority of DLAS have far higher spin temperatures, 
${\rm T_s} > 500$ K (\cite{carilli,ck2000}). Higher ${\rm T_s}$ values are to be expected in 
smaller systems like dwarf galaxies, whose low metallicities and pressures are not 
conducive to the formation the cold phase of HI (\cite{wolfire95}); such systems 
hence have a higher fraction of warm gas as compared to normal spirals, and 
therefore, a high spin temperature. Thus, the 21~cm absorption data appear to 
indicate that the majority of damped systems arise in dwarf or LSB 
galaxies, with only a few systems (at low redshifts) being luminous disks.
Further, deep searches for HI emission from two of these high ${\rm T_s}$ 
objects have resulted in non-detections (\cite{atca,wsrt}), ruling out the 
possibility that they are under-luminous but gas-rich galaxies.

The Chengalur \& Kanekar (2000) compilation of DLAS with published 21~cm absorption
studies has only eighteen systems. We report, 
in this paper, on Giant Metrewave Radio Telescope (GMRT) 21~cm~observations 
of three new candidate and confirmed damped absorbers at low redshift. 
Absorption was detected in two cases, while, for the third system, the 
non-detection imposes a high lower limit on the spin temperature.
The rest of this paper is divided as follows : Sect.~\ref{sec:obs}
describes the GMRT observations. Sect.~\ref{sec:dis} discusses the
derivation of the spin temperatures from our 21 cm observations and also
the properties of 21 cm absorbers in general. All distance dependent quantities
in this paper are quoted for $H_0 = 75$~km~s$^{-1}$~Mpc$^{-1}$ and 
$q_0 = 0.5$.
	
\section{Observations and Data analysis}
\label{sec:obs}
\begin{table*}
\centering
\label{table:OBS}
\caption{Observational details}
\begin{tabular}{ccccccccc}
Source& $z_{\rm abs}$ & $N_{\rm HI}$ & Channel spacing$^\star$ & Total flux & RMS noise &  $\tau$ & $f^\dagger$ & ${\rm T_s}$  \\
& &cm$^{-2}$ &km s$^{-1}$& Jy & mJy & & & K\\
&&&&&&&& \\
PKS~0952+179  & 0.2378 & 2.1e21$^a$ & 2.0 & 1.4 & 2.9  & 0.013      &  0.25  & 2050 $\pm$ 200  \\
B2~0827+243   & 0.5247 & 2.0e20$^a$ & 5.0 &  0.9 & 1.2  & 0.0067     &  0.66 & 330 $\pm$ 70    \\
PKS~0118-272  &0.5579  & 2.0e20$^b$ & 5.0 & 1.1 & 2.4  & $<$ 0.0065$^\ddagger$ &  0.5 & $>$ 850$^\ddagger$         \\
&&&&&&&& \\
\end{tabular}

${}^\star$~ The RMS values are {\it after} Hanning smoothing. \\
${}^\dagger$~ The values quoted are {lower limits} on the covering factor $f$.\\
${}^\ddagger$~Limits are 3$\sigma$.\\
${}^a$~Rao \& Turnshek (2000), ${}^b$~Vladilo et al. (1997)
\end{table*}

The observations were carried out using the GMRT between February and April 2000. A 
30-station FX correlator, which gives a fixed number of 128 channels over a bandwidth 
which can be varied between 64 kHz and 16 MHz, was used as the backend. The number of 
antennas used varied between ten and twelve owing to various debugging and maintenance 
activities; the longest baseline was $\sim 1$ km. Observational details 
are summarised in Table 1; note that the resolution listed is the channel spacing, i.e.
before Hanning smoothing, while the RMS noise is over the smoothed spectrum. We discuss the 
individual sources below, in order of increasing redshift.

PKS~0952+179, $z_{\rm abs} = 0.2378$ : GMRT observations of PKS~0952+179 were carried
out on three occasions, on the 28th of February, and the 3rd and 16th of April, 
2000. A bandwidth of 2 MHz was used on the first of these dates and of 1 MHz on the 
other two (i.e. channel spacings of $\sim$ 4 km s$^{-1}$ and 2 km s$^{-1}$ 
respectively). The total on-source time was 5 hours for the 1 MHz bandwidth observations. 
0839+187 was used for phase calibration and observed every forty minutes in each run; 
3C48 and 3C147 were used for absolute flux and bandpass calibration. 

B2 0827+243, $z_{\rm abs} = 0.5247$ : This source was observed twice, on the 27th of 
March and the 10th of April, 2000, with a bandwidth of 2 MHz (channel spacing $\sim$ 5 
km s$^{-1}$). The total on-source time was three hours in each case. 0851+202 was used 
as the phase calibrator while 3C48 and 3C295 were used to calibrate the system bandpass.

PKS~0118-272, $z_{\rm abs} = 0.5579$ : PKS~0118-272 was observed on the 9th of April, 2000,
using a bandwidth of 2 MHz (channel spacing $\sim$ 5 km s$^{-1}$). 0114-211 was used as
the phase and bandpass calibrator. Bandpass calibration was also carried out using 3C48
and 3C147. The total on-source time was four hours. 

The data were analysed in AIPS using standard procedures. All three target sources 
were unresolved by the GMRT synthesised beam and there was no extended emission present in 
any of the fields. In the case of B2~0827+243 and PKS~0118-272, the field contained a few 
weak compact sources, but quite far from the phase centre; the analysis was hence quite 
straightforward. Continuum emission was subtracted by fitting a linear polynomial 
to the U-V visibilities, using the AIPS task UVLIN. The continuum-subtracted data were then 
mapped in all channels and spectra extracted at the quasar location from the resulting 
three-dimensional data cube. Spectra were also extracted at other locations in the cube 
to ensure that the data were not corrupted by interference.  In case of multi-epoch 
observations of a single source, the spectra were corrected to the heliocentric frame 
outside AIPS and then averaged together.

\begin{figure}
\centering
\psfig{file=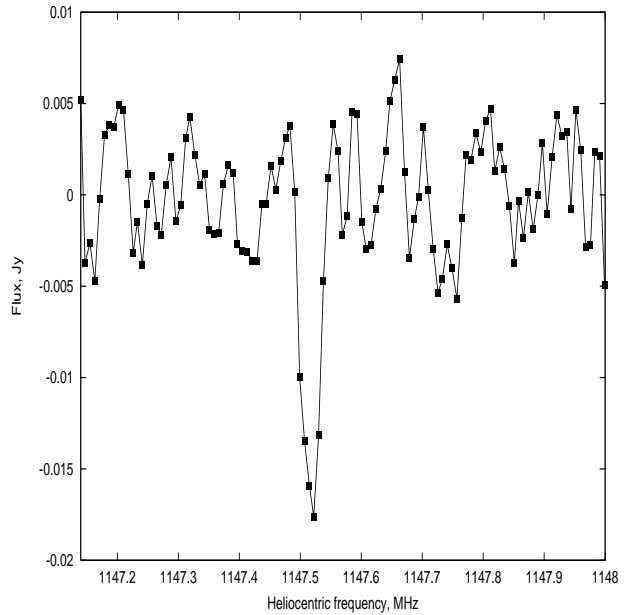,angle=-90,width=3.25truein,height=3.25truein}
\caption{GMRT HI spectrum towards PKS~0952+179. The $x$-axis is
heliocentric frequency, in MHz. The spectrum has been Hanning smoothed and has 
a resolution of $\sim$ 4 km s$^{-1}$.}
\label{fig:0952}
\end{figure}

Fig.~\ref{fig:0952} shows the 1 MHz GMRT spectrum of the absorber towards PKS~0952+179. 
The spectrum has been Hanning smoothed and has an RMS noise level of $\sim$ 2.9 mJy per 
4 km s$^{-1}$ resolution element. Absorption was detected on all three observing runs, with the correct 
Doppler shift. The measured quasar flux is 1.4 Jy; the peak line depth is $\sim$ 18.8 mJy and 
occurs at a heliocentric frequency of 1147.522 MHz, i.e. $z = 0.23780 \pm 0.00002$. The peak 
optical depth is $\sim$ 0.013. 

\begin{figure}
\centering
\psfig{file=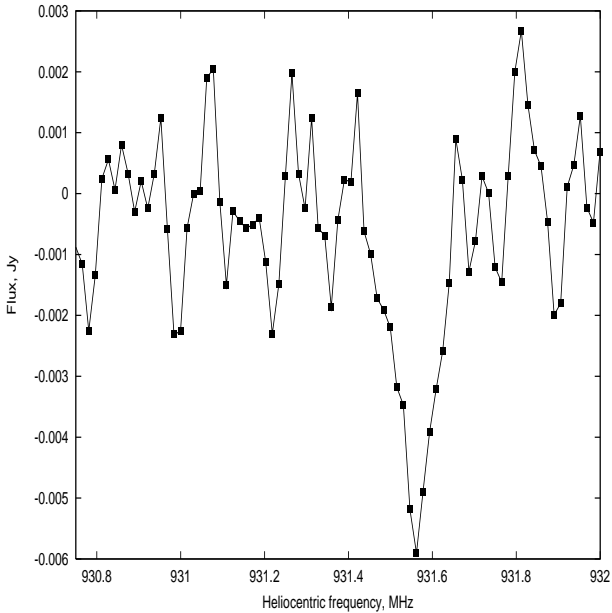,angle=-90,width=3.25truein,height=3.25truein}
\caption{GMRT HI spectrum towards B2~0827+243. The $x$-axis is
heliocentric frequency, in MHz. The spectrum has been Hanning smoothed and 
has a resolution of $\sim$ 10 km s$^{-1}$.}
\label{fig:0827}
\end{figure}

The final Hanning smoothed spectrum of the $z = 0.5247$ absorber towards B2~0827+243 
is shown in Fig.~\ref{fig:0827}; the resolution is $\sim$ 10 km s$^{-1}$. The RMS 
noise on the spectrum is 1.15 mJy while the peak line depth is $\sim$ 6 mJy, i.e. a 
$5.2\sigma$ result. The measured line depth is consistent with the reported non-detection
by Briggs \& Wolfe (1983); their $3\sigma$ upper limit on the line depth was $\sim 7$~mJy.
We note that the absorption was again seen on both observing runs; however, the Doppler 
shift between the two epochs was slightly less than a channel and hence cannot be used 
as a test for the reality of the feature. No evidence for interference was seen in the 
data, on either the source or the calibrators. The absorption is quite wide, with a 
full width between nulls of $\sim$ 50 km s$^{-1}$, and a peak optical depth 
of $\sim 0.0067$, at a frequency of 931.562 MHz, i.e. $z = 0.52476 \pm 0.00005$.

Finally, no absorption was detected in the $z = 0.5579$ absorber towards PKS~0118-272. 
The RMS noise on the final Hanning smoothed spectrum (resolution $\sim 10$ km s$^{-1}$; 
not shown here) is $\sim 2.4$ mJy; this yields a $3\sigma$ upper limit of $\tau < 0.0065$ 
on the optical depth of the absorber. 

\section{Discussion}
\label{sec:dis}
\subsection{The spin temperature}
\label{ssec:Tspin}
\noi The 21~cm optical depth, $\tau_{21}$, of an optically thin, homogeneous cloud is related 
to the column density of the absorbing gas $N_{\rm HI}$ and the spin temperature ${\rm T_s}$ by 
the expression (e.g. \cite{rohlfs86})
\begin{equation}
\label{eqn:tspin}
N_{\rm HI} = { 1.823\times10^{18} {\rm T_s} \over f} \int \tau_{21} \mathrm{d} V \; ,
\end{equation}
\noi where $f$ is the covering factor of the absorber. In the above equation, $N_{\rm HI} $
is in cm$^{-2}$, ${\rm T_s}$ in K and $\mathrm{d}V$ in km s$^{-1}$. For a multi-phase absorber
the spin temperature derived using the above expression is the column density weighted
harmonic mean of the spin temperatures of the individual phases.
In the case of damped systems, 
the column density can be estimated from the equivalent width of the 
Lyman-$\alpha$ profile; a measurement of $\tau_{\rm 21}$ then yields the spin temperature 
{\it if} the covering factor is known. The latter is frequently uncertain since the radio 
emission from quasars is often extended 
while the UV continuum arises essentially from a point source. Thus, the line of sight along 
which the HI column density has been estimated need not be the same as the one for which the 
the 21~cm optical depth has been measured. VLBI observations, when available, can be used 
to estimate the amount of radio emission emanating from compact components spatially 
coincident with the UV point source; one can then estimate $f$ and thus, the spin temperature. 

PKS~0952+179 is a calibrator for the VLA A array, with a flux of 1.1 Jy at 1.4 GHz
(resolution $\sim 1''$), besides some weak extended emission. 2.3 GHz VLBA observations
(see the Radio Reference Frame Image Database of the United States Naval Observatory, 
henceforth RRFID) have resolved this core into a central component and some extended structure, 
with the entire emission contained within $\sim 40$ milli-arcseconds. Of these, the central
component has a flux of $\sim$ 370 mJy at 2.3 GHz, within $\sim 20$ milli-arcseconds; 
this component was further
resolved by VLBA observations at 8.3 GHz (RRFID) into three components, all within 
$\sim 15$ milli-arcseconds. The total emission from these three components is similar
to the flux of the central component at 2.3 GHz, implying that the region
has a fairly flat spectrum. If one assumes that the spectrum of this region remains
flat down to the frequency of our 21~cm observations ($\sim $ 1147.5 MHz) then the small 
size ($\sim 48$~pc at $z=0.2378$) implies that the absorber is likely to cover 
{\it at least} 370 mJy of the total quasar flux. Since the total
flux measured by our GMRT observations is 1.4~Jy, the lower limit to the covering factor
is $f_{\rm min} \sim 0.25$. The column density derived from our observations is then 
$1.02 \pm 0.15 \times 10^{18}\ {\rm T_s} \ (0.25/f)$ cm$^{-2}$. HST observations of the 
Lyman-$\alpha$ line (\cite{rt2000}) give $N_{\rm HI} = 2.1 \times 10^{21}$~cm$^{-2}$, 
comparing this to our 21~cm estimate yields a spin temperature ${\rm T_s} = 2050 (f/0.25) 
\pm 200$~K.  (Note that ${\rm T_s}$ would be higher if a larger amount of 
source flux was covered by the absorber and vice-versa.) The best-fit Gaussian to the 
GMRT spectrum of Fig.~\ref{fig:0952} has an FWHM $\sim 7.7$ km s$^{-1}$. If one 
assumes that this width is entirely due to thermal motions, one obtains a kinetic 
temperature ${\rm T_k} \sim 1500$ K, lower than (but, given the error bars, not very different 
from) our estimate of the spin temperature. Note that standard models of the 
structure of the galactic ISM  (see Kulkarni \& Heiles (1988) for a review) do not 
permit a stable phase with a kinetic temperature of 1500~K;  a detailed analysis of 
heating and cooling mechanisms in the neutral ISM (\cite{wolfire95}) found two stable phases 
(the CNM and WNM), with temperatures in the range 41 - 200 K and 5500 - 8700 K 
respectively (with these values decreasing slightly with a decrease in the metallicity).

In the case of B2~0827+243, the optical quasar spectrum shows strong MgII absorption 
at $z = 0.5247$ (Ulrich \& Owen 1977). HST observations (Rao \& Tunrshek 2000) found 
the Lyman-$\alpha$  line to be damped, but measure a slightly different redshift, 
viz. $z = 0.518$, with $N_{\rm HI} = 2.0 \times 10^{20}$ cm$^{-2}$. No obvious explanation 
exists for the above discrepancy in redshift, except for a possible
error in the wavelength calibration of one of the observations. The GMRT spectrum
of Fig.~\ref{fig:0827} shows 21~cm absorption centred at a redshift of $z = 0.52476$, in 
agreement with the MgII redshift of Ulrich \& Owen (1977); this indicates that the 
HST observations may have had an error in the wavelength correction.

VLA 20~cm A~array observations of B2~0827+243 obtained a flux density of 550~mJy
in an unresolved core component (size $<1''$) (\cite{price93}). This component
was found to remain unresolved in subsequent VLBA observations at 2.3~GHz and
8.5~GHz (resolution $\simeq 20$ milli-arcseconds, this corresponds to a linear
size of $100$~pc at $z=0.5247$) with a flux of $\sim $600 mJy  at both frequencies
(RRFID). Our GMRT observations measured a flux of $\sim 900$~mJy at 931~MHz (in 
good agreement with the 931~MHz flux quoted by Briggs \& Wolfe (1983)); if the 
core contains at least 600~mJy and is completely covered, one obtains  $f \ga 0.66$. 
The column density of the absorbing gas is thus 
$N_{\rm HI} = 6.1 \pm 1.1 \times 10^{18} \ {\rm T_s} \ (0.66/f)$ \cm, from our observations, 
implying a spin temperature ${\rm T_s }= 330 \ (f/0.66) \pm 70$~K. Note that a covering factor of 
unity gives ${\rm T_s} = 470$ K; this is the upper limit on the spin temperature.

The $z = 0.5579$ absorber towards PKS~0118-272 is a candidate 
damped system, for which Vladilo et al. (1997) estimate $N_{\rm HI} \sim 2 \times 
10^{20}$~cm$^{-2}$. This column density is, however, estimated from low-ionization 
metal lines and {\it not} from observations of the Lyman-$\alpha$ transition (see 
also Rao \& Turnshek (2000)). VLBI observations of the quasar at 5 GHz (\cite{shen98}) 
measured 
a flux of 550 mJy within the central 20 milli-arcseconds (corresponding to a linear
size of $100$~pc at $z=0.5579$). Further, a single-baseline 2.3~GHz 
VLBI observation  yielded a flux of $530 \pm 0.05$ mJy (\cite{preston85}). 
The flatness of the core spectrum implies that it is reasonable to expect the core
to contain $\sim 550$ mJy of flux at 911 MHz, our observing frequency. Our GMRT 
observations measured a flux of $\sim 1.1 $ Jy; the lower limit to the covering 
factor is thus $\sim 0.5$. The RMS noise of $\sim 2.4$ mJy (per 10 km s$^{-1}$ channel) 
then gives a 3~$\sigma$ lower limit of $850 (f/0.5)$ K on the spin temperature.

Deep K-band imaging and spectroscopy of the B2~0827+243 field (\cite{raovla})
have identified a spiral galaxy at $z = 0.524$ as the system giving rise to the damped 
absorption. The low estimated spin temperature of the absorber continues the trend
noted earlier (\cite{ck2000}) for DLAS associated with spirals to have low ${\rm T_s}$.
Interestingly, the impact parameter of the quasar line of sight to the absorbing galaxy 
is $\sim$ 28 kpc (\cite{raovla}); the low spin temperature then implies that substantial
amounts of cold HI are present even at such large distances from the galaxy centre. 

	The field around PKS~0952+179 has also been imaged by Turnshek et al. (2000) 
in the K-band, and  the absorber tentatively identified as an LSB galaxy. No large 
spiral is present in the field, consistent with the high spin temperature obtained
for this system.  For the candidate DLAS PKS~0118-272 Vladilo et al. (1997) find that
the [Ti/Fe] and [Mn/Fe] ratios are similar to that of diffuse ISM clouds and identify
by PSF subtraction an candidate absorbing galaxy at an impact parameter of 8~kpc;
the redshift of this galaxy is however unknown. If the galaxy is at the same redshift
as the absorption line system it would have $M_R \sim -21$. 

\subsection{ Properties of 21 cm absorbers }

\begin{figure}
\centering
\psfig{file=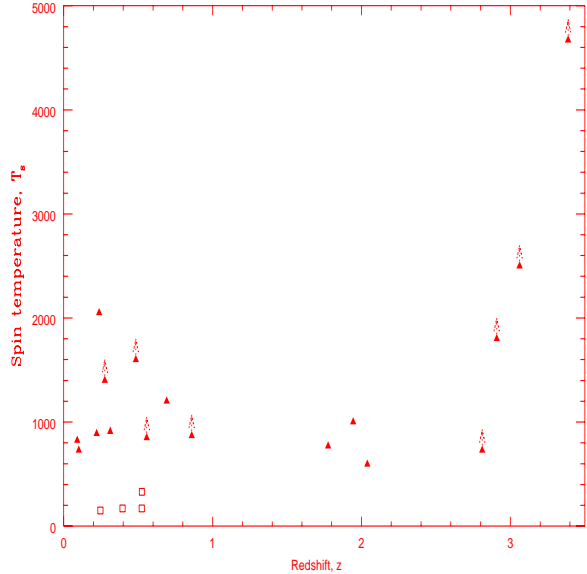,angle=-90,width=3.25truein,height=3.25truein}
\caption{ The figure shows the spin temperature ${\rm T_s}$ plotted against redshift, 
for 21 DLAS, excluding the $z=0.437$ absorber towards 3C196. DLAS identified as 
spiral galaxies are shown as hollow squares, while lower limits on the spin 
temperature are indicated by upward arrows.}
\label{fig:tvsz}
\end{figure}

\begin{figure}
\centering
\psfig{file=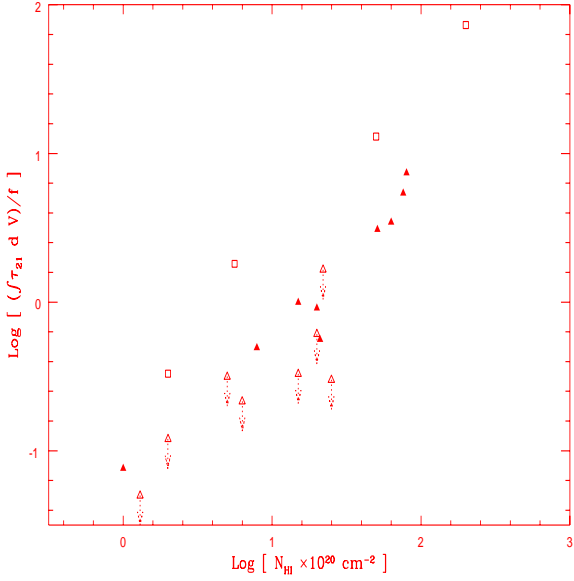,angle=-90,width=3.25truein,height=3.25truein}
\caption{ The figure shows the ``true'' 21 cm equivalent width $EW = [\int \tau \mathrm{d}V]/f$
plotted against HI column density $N_{\rm HI}$, on a logarithmic scale. Spiral galaxies
are indicated by hollow squares, while filled and open triangles indicate detections
and non-detections of absorption, respectively, in the remaining systems.}
\label{fig:eq_wid}
\end{figure}

The sample of DLAS with published 21~cm observation observations 
presently consists of 22 systems (the 18 absorbers listed in (\cite{ck2000}), 
the three described here and the $z = 0.101$ candidate towards PKS~0439-433 
(\cite{atca})). Fig.~\ref{fig:tvsz} shows a plot of spin temperature versus 
redshift, for 21 systems of the sample (the $z \sim 0.437$ absorber towards 3C196 is 
excluded since its $N_{\rm HI}$ is uncertain); it can be seen that the majority of 
DLAS have ${\rm T_s}$ values far higher than those of the Milky Way, with the only 
exceptions being systems identified as spiral galaxies. If damped systems have 
the same two-phase structure seen in our Galaxy, this would mean that 
the absorbers have higher amounts of the warm phase of neutral hydrogen (WNM) than 
normal spirals (\cite{ck2000}). Fig.~\ref{fig:eq_wid} plots the ``true'' 
21 cm equivalent width (i.e. $EW = [\int \tau_{21} \mathrm{d} V]/f$) of 21 systems 
of the sample (again excluding the $z=0.437$ DLA towards 3C196) against HI column density, 
on a logarithmic scale; it can be seen that the trend for increasing $EW$ with 
$N_{\rm HI}$, seen in Galactic data (\cite{carilli}), persists for \dla systems. 
More importantly, DLAS identified as 
spiral galaxies (open squares) can be seen to have systematically higher equivalent 
widths, at the same values of HI column density. Further, most DLAS (except the ones 
identified as spirals) tend to lie around the lower edge of the envelope defined 
by the Galactic data (see Carilli et al. (1996) for a comparison). This also 
indicates (as suggested earlier by Carilli et al. (1996)) that the majority of DLAS 
contain a higher fraction of the warm phase of neutral hydrogen than normal spirals 
like the Milky Way, and are hence unlikely to be spiral galaxies. Similarly, Petitjean 
et al. (2000)) used observations of molecular hydrogen in a sample of DLAS 
to conclude that most damped absorption arises in warm ($T > 3000$~K) gas.

Direct evidence for the proposition that damped systems have a larger WNM content as
compared to local spirals comes from the high resolution Arecibo observations 
of the two DLAS towards OI363 (\cite{lane2000,wnm}). In the case 
of the lower redshift ($z = 0.0912$) system, the ratio of column densities of 
the WNM to CNM is $\ga $ 2:1 (\cite{lane2000}), while, for the $z = 0.2212$ absorber, 
this ratio is $\ga $ 3:1 (\cite{wnm}). Thus, at least in these two systems, the 
WNM to CNM ratio is higher than that seen in spiral galaxies and is, in fact, 
similar to values obtained in dwarf systems (\cite{young2000}).

\begin{figure}
\centering
\psfig{file=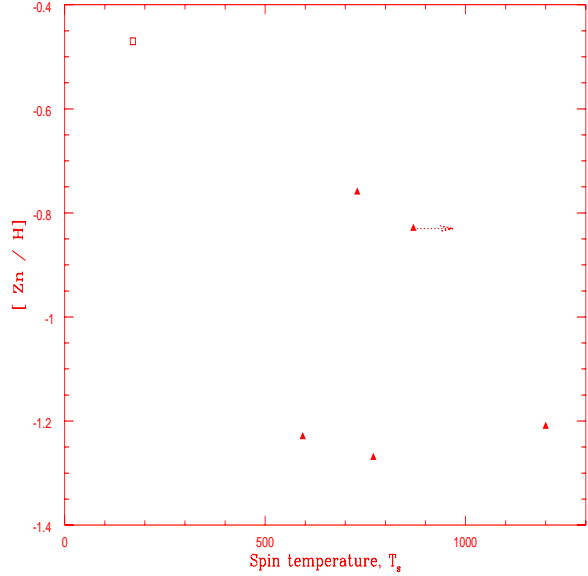,angle=-90,width=3.25truein,height=3.25truein}
\caption{ The figure shows the [Zn/H] ratio plotted against spin temperature
 ${\rm T_s}$,
for the six damped systems with [Zn/H] estimates. The $z \sim 0.395$ absorber towards
PKS~1229-021 is indicated by a hollow square.}
\label{fig:znts}
\end{figure}

        If the high spin temperatures of the majority of damped absorbers
arise due to their lower metallicities and hence, larger fractions of the WNM 
(\cite{ck2000}), one would expect that the metallicities (i.e. the [Zn/H] ratios) 
of the DLAS identified as spirals should be systematically higher than those 
of systems with high ${\rm T_s}$ values. Unfortunately, [Zn/H] estimates exist 
for only six DLAS of the sample and for only one low ${\rm T_s}$ system (the $z \sim 0.395$ 
absorber towards PKS~1229-021) (\cite{boisse98,pettini94,pettini97,meyer89,meyer92}). 
These values are plotted against spin temperature in Fig.~\ref{fig:znts}; it can be 
seen that the low ${\rm T_s}$ system has the highest [Zn/H] ratio of all six absorbers. 
It would be very interesting to test whether this trend persists, for the other 
absorbers in the sample.

\begin{figure}
\centering
\psfig{file=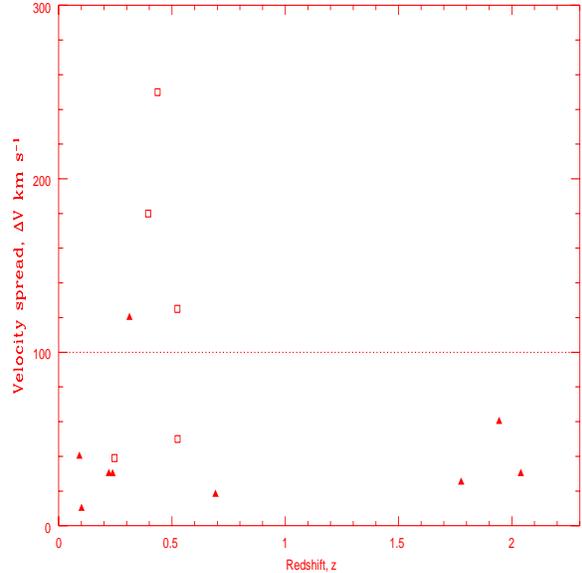,angle=-90,width=3.25truein,height=3.25truein}
\caption{ The figure shows the 21~cm velocity spread $\Delta V$ plotted against redshift
for all damped systems with detected 21~cm absorption. Absorbers identified as
spiral galaxies are shown as hollow squares.}
\label{fig:vel}
\end{figure}

\begin{figure}
\centering
\psfig{file=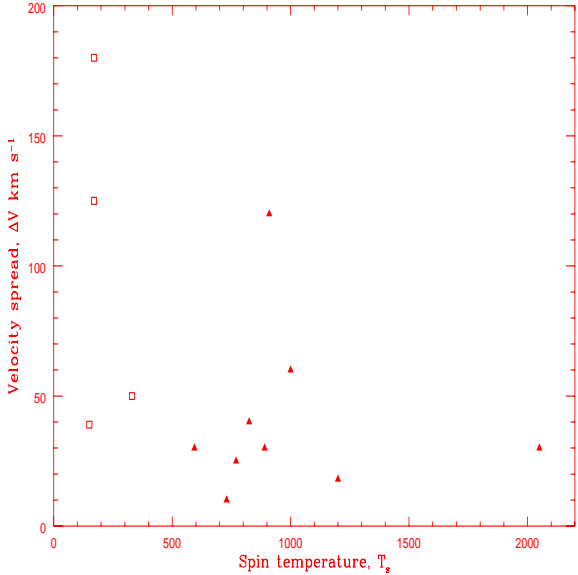,angle=-90,width=3.25truein,height=3.25truein}
\caption{The figure shows the 21~cm velocity spread $\Delta V$ plotted against spin
temperature ${\rm T_s}$, for all damped systems with detected 21~cm absorption. Absorbers
identified as spiral galaxies are shown as hollow squares. Note that the $z = 0.437$
absorber towards 3C196 has not been included in the figure since its spin temperature
is unknown; see discussion in text.}
\label{fig:vel_ts}
\end{figure}

Figs.~\ref{fig:vel} and \ref{fig:vel_ts} plot the velocity
spread $\dV$ (full width between nulls) of the systems with detected 21~cm 
absorption against redshift and spin temperature, respectively.  (The total velocity 
spread is used, instead of the FWHM of the absorption profile, to account for the 
possibility that a single line of sight intersects multiple clouds, as is typical 
in spiral galaxies.) The plot of $\dV$ versus redshift indicates
that large velocity spreads ($\Delta V > 100$ km s$^{-1}$) are not found for $z
\ga 2$, while both large and small spreads are seen at low redshift. Similarly, 
Fig.~\ref{fig:tvsz} shows that, while both low and high ${\rm T_s}$ values are obtained 
at low redshift, only high values are obtained for $z \ga 1.5$. This is qualitatively 
consistent with what is expected in hierarchical clustering models (e.g. Kauffmann 1996), 
 in which systems with low circular velocities dominate the absorption 
cross-section at high redshift, while systems with both high and low circular 
velocities are found at low $z$. Next, Fig.~\ref{fig:vel_ts} shows a possible
correlation between the velocity width and the spin temperature of the absorbers,
with only one system (the $z = 0.3127$ absorber towards PKS~1127-145) having
both a large 21~cm velocity spread and a high temperature; all other systems with large
($\dV > 100$ km s$^{-1}$) velocity widths are identified with low ${\rm T_s}$ spirals.
The system towards PKS~1127-145 is likely to be tidally disturbed 
since there are at least three galaxies at the redshift of the \dla absorber 
(\cite{lane98}). (Note also that the $z \sim 0.437$ absorber towards 3C196 has not 
been included in this figure, since its ${\rm T_s}$ is unknown. However, the observations 
are consistent with ${\rm T_s} \la 200$~K (\cite{debruyn2000}), while the velocity spread 
is 250 km s$^{-1}$; it would hence lie close to the upper left corner of the figure.)
The average velocity spread for the absorbers identified as spirals 
is $\dV_{ave} \sim 130$ km s$^{-1}$ while, for the high ${\rm T_s}$ systems, the mean 
spread is far lower, $\dV_{\rm ave} \sim 40$ km s$^{-1}$. 

It should be emphasised that the results on the velocity widths of the absorption
lines stem from the sub-sample of 14 systems which show 21~cm absorption; the
statistics should clearly be improved before any strong conclusions can be drawn.
However, although the numbers are small, the general trends in ${\rm T_s}$ and $\Delta V$ are
consistent with scenarios in which \dla absorbers are typically small systems at
high redshift, while, at low redshift, they are a composite population including
both spiral galaxies as well as smaller systems. 

\begin{acknowledgements} These observations would not have been possible without the
many years of dedicated effort put in by the GMRT staff in order to build the telescope.
This research has made use of the United States Naval Observatory (USNO) Radio 
Reference Frame Image Database (RRFID). 
\end{acknowledgements}

\end{document}